\documentclass{article}
\usepackage{amssymb}
\usepackage{amsmath}
\usepackage{graphicx}

\begin{document}

\title{From Fractional Quantum Mechanics to Quantum Cosmology: An Overture}
\author{Paulo Vargas Moniz$^{1}$\thanks{ pmoniz@ubi.pt} and Shahram Jalalzadeh$^{2}$\thanks{shahram@df.ufpe.br}
\\
\small$^{1}$Departamento
de F\'isica and Centro de Matem\'atica e Aplica\c c\~oes (CMA-UBI), \\ \small Universidade da Beira Interior,  Rua Marqu\^es d\'Avila e Bolama, 
  \\\small 6200 Covilh\~a,  Portugal\\
\small$^{2}$Departmento de F\'{i}sica, Universidade Federal de Pernambuco, Pernambuco,\\\small PE, 52171-900, Brazil }
\date{\today}
\maketitle

\begin{abstract}
 Fractional calculus is a couple of centuries old, but its development 
has been less embraced and it was  only within the last century that a program of applications for physics started. Regarding quantum physics, it has been only in the previous decade or so that the corresponding literature resulted in a set of defying papers. In such a context, this manuscript constitutes a cordial invitation, whose purpose   is simply to suggest, mostly through a heuristic and unpretentious presentation, the extension of fractional quantum mechanics to cosmological settings.   Being more specific, we start by outlining a historical  summary of fractional calculus. 
 Then, following this motivation, a (very) brief appraisal of fractional quantum mechanics is presented, but where  details (namely those of a mathematical nature) are  left  for literature perusing. 
Subsequently, the application of fractional calculus  in  quantum cosmology 
is  introduced, advocating it as worthy to consider: if the progress of fractional calculus serves as argument, indeed   useful consequences will also be drawn (to cite from Leibnitz).
In particular, we discuss different difficulties that may affect the operational framework to employ, namely the issues of minisuperspace covariance and fractional derivatives, for instance. An example of investigation is provided by means of a very simple model. Concretely, we restrict ourselves  to speculate that with  minimal  fractional calculus  elements, we may have a peculiar  tool to inspect the flatness problem of standard cosmology. In summary, the subject of fractional quantum cosmology is    herewith proposed, 
merely realised in terms of an open program constituted by several challenges.

    
\end{abstract}

\section{ Historical (Introduction)}
\label{sec-1}


Fractional  calculus  follows from a  question~\cite{A}:
Can  the  meaning  of  derivatives  (of  integral
order $\left. \frac{d^n y}{dx^n} \right)$  be  extended  to  have  the case  where  $n$  is  any  number, i.e.,
irrational,  fractional  or  complex?


L'Hospital    asked  Leibnitz (Leibnitz  invented  the  above  notation)  about  the
possibility  that  $n$  be  a  fraction, who, delphically,  then  suggested
``($\dots$)  useful  consequences  will  be  drawn.'' As if complying to the oracle,
Lacroix  later advocated  the  formula ($\Gamma$  is Legendre's  symbol,  a~ generalized  factorial)
\begin{equation}
    \frac{d^\frac{1}{2} y}{d x^\frac{1}{2}}  =
    \frac{\Gamma (a+1)}{\Gamma \left(a+\frac{1}{2}\right)}  x^{a-\frac{1}{2}},
    \label{1}
\end{equation}
which  expresses  the  derivative  of   order  $1/2$  of  the
function  $x^a$. For~ $y  =  x$
\begin{equation}
   \frac{d^\frac{1}{2} x}{d x^\frac{1}{2}}  =
    \frac{2\sqrt{x}}{\sqrt{\pi}}.
    \label{1a}
\end{equation}
Abel  applied   fractional  calculus to the  tautochrone  problem~\cite{A},
whose elegant solution  enthused  Liouville. Riemann    while  a  student set the path
to the  present  day  Riemann-Liouville  definition  of  a  fractional  derivative~\cite{A}.

Nonetheless, fractional calculus is not yet generally known.
The challenge is to establish results, serving as
justifications,  so as to lead and  popularize  the  topic. This would, hopefully,
 further enthuse  scientists    to  either explore or apply  it  into  their  research.
Fractional  calculus has assisted in rheology,  quantitative  biology,
electrochemistry,  scattering  theory,  diffusion,  transport  theory,
probability,  potential  theory  and  elasticity~\cite{A}.
  Thus,  whereas  the  theory
of  fractional  calculus  has  been developing,  its subsequent  use  needs
encouragement, specifically towards   physical  phenomena
 that  can  be  treated  with  the  elegance  of
fractional  calculus~\cite{Herrmann,Herrmann-n}.

Therefore, it was only sensible to embrace fractional calculus and explore it within  quantum mechanics, which has led  to very interesting features 
indeed (we mention the possibility of relating fractal features to fractional (quantum) mechanics, see~\cite{B} and references therein)
cf.~\cite{B,B-n,B-fqm,B-fqm-n,B-fqm-nn}, see also~\cite{T1,T2}. As~we will  briefly point out, a~generalized path 
integral lays importantly at the essence of  fractional quantum mechanics~\cite{B}.

On the other hand, it has been  established how the Wheeler--DeWitt equation, a~paradigmatic tool in quantum cosmology, can be assembled  from the Brownian--Feynman path integral~\cite{C1,C2,NEWqm} 
 So, could that  procedure (the generalized path integral, central in fractional quantum mechanics)  be extended towards a fractional (minisuperspace) quantum cosmology  set-up? 
 What would be the obstacles to address?  Should  heuristic insights be taken aboard, providing complementary targets to investigate? Trustfully,  importing from Leibnitz's omen~\cite{A},   useful  consequences  would  be  drawn, whatever the conclusions to be~extracted.

The paper is organized as follows. In~Section~\ref{sec-1} a very brief   historical  summary of fractional calculus  is presented; an outstanding review can be found in~\cite{A}, but~we also suggest~\cite{B}.  Fractional quantum mechanics (and calculus) is unveiled in Section~\ref{sec-2}, constituting now a subject with  a vast domain and whose literature is getting  wider; for further  technical aspects,  we  suggest  the  works~\cite{B,B-fqm} indicated in the bibliography.
Then, in~Section~\ref{sec-3}, we describe a few features of quantum cosmology and path integral formalism~\cite{C1,C2}, in~particular discussing them within the scope of 
a (general) path integral, that intrinsically assists fractional quantum mechanics~\cite{B,B-fqm,T1,T2}. 
In Section~\ref{sect-4} we speculate on  the application of fractional calculus  in  quantum cosmology.   An~example  is heuristically provided, whereby we only consider (as application)  the flatness problem of standard cosmology. Finally, in~Section~\ref{sec-5} we conclude the work and speculate on future challenges to be addressed.


\section{ Fractional Quantum~Mechanics}
\label{sec-2}


Canonically,
the Hamiltonian function has the form
\begin{equation}
    H(\mathbf{p}, \mathbf{r}) :=
\frac{\mathbf{p}^2}{2m}
+ V (\mathbf{r}),
\label{2}
\end{equation}
where $\mathbf{p}$ and $\mathbf{r}$ are, respectively,
the momentum and space coordinate of a particle with
mass $m$ and $V (\mathbf{r})$ is the potential energy. Quantum mechanically,
$\mathbf{p}$ and $\mathbf{r}$ become
 operators $\hat{\mathbf{p}}$ and $\hat{\mathbf{r}}$ and  the Hamiltonian
proceeds towards
\begin{equation}
   \hat{H}(\hat{\mathbf{p}}, \hat{\mathbf{r}}) :=
\frac{\hat{\mathbf{p}}^2}{2m}
+ \hat{V} (\hat{\mathbf{r}}),
\label{2a}
\end{equation}
where $\hat{V} (\hat{\mathbf{r}})$ is the potential energy~operator.

As in the previous section, let us import another quite `unexpected'  question: are there other forms (for the kinematic term in Equations~(\ref{2}) and  (\ref{2a})) which do not contradict the fundamental principles of classical mechanics and quantum
mechanics~\cite{B}?

In essence, addressing that challenge has been achieved and
let us present a very succinct  summary. Being more concrete, 
fractional quantum mechanics emerged from a generalized path integral framework, from~which a (generalized) fractional Schr\"odinger equation can extracted
~\cite{B,B-fqm,T1,T2}.
So, as~far as the current approach to  fractional quantum mechanics is concerned, 
it is necessary to consider two stages. 
On the one hand, to~widen the tooling range  from the straight, albeit useful, canonical methodology,
towards the language   of the path integral. The~canonical representation and the path integral description are inter-related~\cite{NEWqm}. For~instance, the~Schr\"odinger equation follows from either. Nevertheless,
the path integral  is far wider as operational application (allowing to sum  different paths;  
for example,  different  geometries within distinctive  topological classes, concerning quantum cosmology). On~the other hand,   to~navigate the path integral within 
fractional calculus, we need to employ the  larger context of L\'evy~paths.


\subsection{L\'evy~Paths}
\label{subsect-levy}

L\'evy and Brownian paths (the latter is a particular case of the former) are associated with  stochastic (or Wiener) processes, with~segment-like motion  proceeding between spatial points, described from  a few mathematical assumptions. Namely, some degree of continuity (for the Brownian process) or not at all:  Brownian motion has continuous paths, whereas others (fitting within the wide L\'evy scope) may not. The~admission of `jumps' in the wider L\'evy (and not Brownian) context for paths, has been of interest in exploring, namely in  quantum physics~\cite{B}. 
If  the reader is interested, please consult~\cite{B,B-fqm,Herrmann,NEW,NEW-n,NEW-nn} and references~therein. 

A brief selection of a few particulars follows~\cite{NEW}:


\begin{itemize}
    \item Feynman's path
integral operates over Brownian-like paths. Nevertheless,
 Brownian motion is a special
case of  
$\alpha$-stable (In probability theory, a~distribution is said to be stable if a linear combination of two independent random variables with this distribution has the same distribution;   please  see~\cite{NEW}) 
probability distributions;
\begin{itemize}
    \item Will the sum of $N$ independent identically distributed random quantities
    $X = X_1 + X_2 + \cdots + X_N$ have
the same probability distribution   as each single
 $p_i (X_i), i = 1, $\dots$N$ ?
\item Each $p_i (X_i)$ proceeds to be a Gaussian
(cf. central limit theorem);
\item Furthermore, a~sum of $N$
Gaussian functions is again a Gaussian.
\end{itemize}
\item However,  there exist the possibility to generalize the
central limit theorem;
\begin{itemize}
    \item There is  a class of non-Gaussian
    $\alpha$-stable
     probability distributions,  bearing a
 parameter
$\alpha$, designated as L{\'e}vy index, with~range as  $0 < \alpha \leq 2$;
\item When $\alpha = 2$, we recover  Brownian
motion (If the fractal dimension~\cite{B}  of the Brownian path is $d_{fractal}
=2$, then the L{\'e}vy motion  has fractal dimension $d = \alpha$, where $\alpha$ now
$1 < \alpha \leq 2$)
\end{itemize}
\end{itemize}

Therefore, the~L{\'e}vy index $\alpha$ would become a  fundamental parameter in (fractional) classical and quantum
 mechanics. And~with a distinction between the (fractal)
dimensions of the Brownian and L{\'e}vy paths~\cite{B},  that
would imply significant differences concerning
the behaviour of physical 
systems.

Let us mention, also briefly, that  having been pursued within applied mathematics domains,   fractional quantum mechanics has not been systematically explored  with a view towards  laboratory experiments. Nevertheless, discussions and papers have emerged; references~\cite{Pin,Pin-a} constitute a sample from the literature, although~not reporting actual  work, \textit{directly} involving fractional quantum features.
Specifically, in~\cite{Pin},   solid state physics was regarded, involving  the effective mass $m(k)$, in~concrete Bose-Einstein condensate systems. To~the best of our knowledge, 
virtually no concrete observational or experimental progress has been  attempted; solely theoretical features and a few quantities  with specific formulae 
or ranges were computed.  Fractional quantum mechanics has not yet been tested, but~it is  falsifiable plus consistent,  in~that includes standard quantum mechanics (as clear limiting cases through parameter variation). 

\subsection{(Quantum) Mechanics}


The Hamiltonian function specifically becomes $H_\alpha
(\mathbf{p}, \mathbf{r})$,
as
\begin{equation}
    H_\alpha
(\mathbf{p}, \mathbf{r}) : = D_\alpha
|\mathbf{p}|^\alpha
+ V (\mathbf{r}), \, \,  1 < \alpha \leq  2,
\label{5}
\end{equation}
with  $D_\alpha$
being a coefficient.
We stress that   L\'evy path integrals allow to generalize  standard
quantum mechanics, based on the well-known Feynman path integral: the latter yields the  Schr\"odinger
equation, whereas the former (over L\'evy trajectories) leads to the corresponding fractional Schr\"odinger~equation.

Therefore, 
let us  just unveil
that the fractional Schr\"odinger equation will include a derivative over spatial coordinates
but of order $\alpha$, 
instead of the usual second  order space 
derivative.

The operators are introduced as follows,
\begin{equation}
    E \rightarrow  i\hbar \frac{\partial }{\partial  t}, \, \,
    \mathbf{p} \rightarrow  -i\hbar \mathbf{\nabla},
    \label{6}
\end{equation}
with, as~usual,
$\nabla = \frac{\partial }{\partial  \mathbf{r}}$
and $\hbar$ being  Planck's constant over $2\pi$. The~fractional
Schr\"odinger
equation is written as
\begin{equation}
    i\hbar \frac{\partial \psi(\mathbf{r},t)}{\partial  t}
= \hat{H}_\alpha(\hat{\mathbf{p}}, \hat{\mathbf{r}}) \psi(\mathbf{r},t)
:=
D_\alpha
(-\hbar^2 \Delta)^{\alpha/2} \psi(\mathbf{r},t)
+ V(\mathbf{r},t) \psi(\mathbf{r},t),
\label{7}
\end{equation}
with $1 < \alpha \leq 2$ and  $(-\hbar^2 \Delta)^{\alpha/2}$ being a
generalization of the fractional (quantum) Riesz derivative~\cite{B},
written as
\begin{equation}
(-\hbar^2 \Delta)^{\alpha/2} \psi(\mathbf{r},t)
=
\frac{1}{(2\pi\hbar)^3}
\int d^3 p e^{i\frac{\mathbf{p}\cdot \mathbf{r}}{\hbar}}
|\mathbf{p}|^\alpha
\varphi(\mathbf{p},t),
\label{8}
\end{equation}
by means of Fourier transforms, to~relate
$\psi(\mathbf{r}, t)$ and   $\varphi(\mathbf{p}, t)$;  $\Delta$ is the Laplacian.
For the special case when $\alpha = 2$ and $D_2
= 1/2m$, where $m$ is the particle
mass (Extracting from~\cite{CKTP}, in~Brownian-like motion a diffusion constant $D$ is associated,
proportional to $\hbar$, as~$\hbar=D M$, $M$ with mass dimensions, varying from case (i.e., particle) to case. $M$ can be matched experimentally with good accuracy to the inertial mass; the inertial mass (equal to the gravitational mass) would thus be associated with the `quantum' mass and both originating from  energy momentum tensor emerging in the Wheeler--DeWitt equation),
we recover the standard  Schr\"odinger~equation.

Before proceeding, let us mention a  pertinent aspect within fractional calculus. From~a  purely mathematical point of view, the~use of dimensions and hence of homogeneity within formulae with dimensional quantities (physical observables) is meaningless. However,~this may be different if   proceeding eventually towards equations for a physical system to be tested. 
This issue could become of importance when bringing fractional quantum mechanics (and cosmology) towards realistic experiments. It would be therefore  of relevance to investigate  issues of dimensionality arising from fractional derivatives; if (and how), they 
 could become  hidden  in the constants, taken as 
parameters to fit. Cf. e.g.,~\cite{Cresson,Inizan}.

\subsection{The Case of $H_\alpha =0$}


Let us very briefly comment on the special case
~\cite{B}
when the Hamiltonian $H_\alpha$
does not depend explicitly on the
time (Although the content in this subsection is entirely non-relativistic (see~\cite{B}), this  case study  is of interest (strictly in formal terms, we emphasize) in quantum cosmology, whereby the Wheeler--DeWitt equation also bears a $H=0$ character, albeit quite different in context and meaning)
Accordingly,
there exist the  solution
 of the form (we take the one-dimensional case for ease of notation)
\begin{equation}
    \psi(x, t) = \exp{ \left( - \frac{iEt}{\hbar}
    \right) }
\phi(x), \label{8a}
\end{equation}
where $\phi(x)$ satisfies (please cf.~\cite{B} for details)
\begin{equation}
    H_\alpha \phi(x) := -D_\alpha(\hbar \nabla)^\alpha \phi(x) + (V(x) - E) \phi(x)
    =    0, \label{9}
\end{equation}
with, recalling,  $1 < \alpha \leq 2$.

Equation~(\ref{9}) is the time-independent  fractional
Schr\"odinger equation. Likewise, we could speculate and
assign, in~this `fractional' context, the~probability to find a particle at
 $x$ as  the absolute square of the
wave function $|\psi|^2$ or  $|\phi|^2$, as~ above.

\subsection{Harmonic Oscillator and~Beyond}\label{HO}


A physical application of traditional fruitfulness
is with a potential given by~\cite{B}
\begin{equation}
V (|{\mathbf{r}_i} -  {\mathbf{r}_j}|)
\simeq
|{\mathbf{r}_i} -  {\mathbf{r}_j}|^\beta,
    \label{11}
\end{equation}
with $\beta > 0$.

The corresponding fractional Hamiltonian operator $H_{\alpha, \beta}$
is provided  as
\begin{equation}
H_{\alpha, \beta} = D_\alpha
(-\hbar^2 \Delta)^{\alpha/2} + |{\mathbf{r}}|^\beta.
    \label{12}
\end{equation}
For the special case, when $\alpha = \beta$, assuming $1 < \alpha \leq 2$,
the Hamiltonian can be considered as 
the fractional generalization of
the harmonic oscillator Hamiltonian of standard quantum~mechanics.

The one-dimensional fractional oscillator~\cite{B} provides
pertinent  semiclassical features. Setting
$E \equiv D_\alpha |p|^\alpha + |x|^\beta$, remembering
that $|p| =0$ at the turning points, the~standard
Bohr-Sommerfeld quantization rule
instructs to 
take
\begin{equation}
2\pi\hbar\left( n + \frac{1}{2} \right)
= \oint p dx = 4 \int_0^{x_m} p dx =
4 \int_0^{x_m} D_\alpha^{-1/\alpha} (E - |x|^\beta)^{1/\alpha} dx,
    \label{13a}
\end{equation}
where $\oint$ indicates the integral over one complete period of the
classical motion; $x_m \sim  E^{1/\beta}$
is the turning point of classical motion.  There are turning points at $|x|=x_m$ and the integral in (\ref{13a}) is from $0$ to  $x_m$, not in between the turning points. The~latter would  make a factor of $2$ to be used 
but in (\ref{13a}),  a~different description was clearly  taken; please see~\cite{B,NEWqm}.

The energy can be presented as
\begin{equation}
E_n = \left(
\frac{\pi \hbar \beta D_\alpha^{1/\alpha}}{2B
\left( \frac{1}{\beta},
{\frac{1}{\alpha}+1}\right)}
\right)^\frac{\alpha\beta}{\alpha+\beta}
\left(n+\frac{1}{2}\right)^\frac{\alpha\beta}{\alpha+\beta}
    \label{13b}
\end{equation}
and for $\alpha=\beta=2$ we recover the
result (The B-function is defined by
$B(u, v) = \int_0^1 dy y^{u-1}
(1 - y)^{v-1}$) of
the standard
quantum mechanical oscillator. It is curious to emphasize that
for
\begin{equation}
\frac{1}{\alpha} + \frac{1}{\beta} = 1,
    \label{13c}
\end{equation}
the spectrum is equidistant, and~that when assuming
$1 < \alpha, \beta \leq 2$,  that is only allowed for
$ \alpha = \beta = 2$.

\subsection{Tunneling}


The tunneling of a particle is a paradigmatic feature of quantum mechanics. The~tunneling problem within fractional quantum mechanics has been solved for various potential configurations (cf.~\cite{T1,T2}
and references therein). Interestingly, the~Hartman
effect (concretely, the~tunneling time being independent of the width of the barrier for sufficient thickness) seems non-existent in  fractional quantum mechanics~\cite{B,T1,T2}.
In particular, for~a square barrier with potential $V(x)=V$ ($V$ a constant) confined to $0\leq x \leq b$ and zero elsewhere, the~general solution of the corresponding  fractional Schr\"odinger equation is
\begin{equation}
\psi(x) = \left\{
\begin{array}{ll}
    A e^{ik_\alpha } + B A e^{-ik_\alpha }, & x < 0, \\
    C \cos \overline{k}_\alpha  + D \sin \overline{k}_\alpha, & 0 < x < b, \\
    F e^{ik_\alpha } + G e^{-ik_\alpha }, & x > b,
\end{array}
\right.
    \label{tunnel-1}
\end{equation}
where
\begin{equation}
  k_\alpha = \left( \frac{E}{D_\alpha \hbar^\alpha} \right)^\frac{1}{\alpha}
    \label{tunnel-1a}
\end{equation}
and
\begin{equation}
  \overline{k}_\alpha = \left( \frac{E-V}{D_\alpha \hbar^\alpha} \right)^\frac{1}{\alpha}.
    \label{tunnel-1b}
\end{equation}
From (\ref{tunnel-1})--(\ref{tunnel-1b}) (or the explicit expressions associated with other potentials and cases) we can extract, e.g.,~transmission coefficients, depending on $\alpha$  
\cite{T1,T2}.

The essential feature to bear in mind is that in fractional quantum mechanics the path integral is taken over L\'evy paths, meaning a higher probability for particles to travel farther per `jump' in contrast to Brownian--Feynman paths~\cite{B,B-fqm,T1,T2}.  This emerges from the fact that L\'evy paths are generalizations of Brownian-segments, meaning that they account for probability distributions, allowing infinite variance and inducing a non-negligible probability to reach far away points over a longer step, in~comparison to the standard ones from Brownian--Feynman's (cf.~\cite{B,B-fqm,T1,T2,NEW}).

Another interesting feature is retrieved in the case of delta and double-delta~\cite{T2} potential: There is tunneling,  even at zero energy. This comes from the application of the uncertainty principle, which in fractional quantum mechanics is~\cite{T2}
\begin{equation}
\langle | \Delta x |^\mu \rangle^\frac{1}{\mu}
\langle | \Delta p_x |^\mu \rangle^\frac{1}{\mu}
>
\frac{\hbar}{(2\alpha)^\frac{1}{\mu}},
    \label{unc-prin}
\end{equation}
with $\mu < \alpha $, $1< \alpha \leq 2$; the standard quantum mechanics expression
is recovered for $\mu = \alpha = 2$. It should then be noticed that for $E=0$, we can have energies as
\begin{equation}
\Delta E \sim \frac{\langle | \Delta p |^\mu \rangle^\frac{2}{\mu}}{2m} ,
    \label{unc-prin-a}
\end{equation}
with momentum $\langle | \Delta p |^\mu \rangle^\frac{1}{\mu} $.

\section{Quantum Cosmology and (General)   Path Integral}
\label{sec-3}

Generically, a~relationship between the canonical (specifically, Dirac-like)
and path-integral quantization 
was discussed in~\cite{C1,C2} for minisuperspace models (i.e.,
quantum cosmology). Merely extracting 
and summarizing the essential guideline from the abstract 
in~\cite{C1}, let us add
the main point. 
It was shown that the path-integral 
framework allowed to obtain expressions, 
that were shown to satisfy the constraints, namely the Wheeler--DeWitt equation.
Notwithstanding the significant and fundamental contribution from~\cite{C1}, a~derivation
of the Wheeler--DeWitt equation in full quantum gravity was not given, either there or 
elsewhere~\cite{C1,C2}. Moreover,  criticisms and appraisals were raised for other (strictly formal) approaches and papers therein cited, concerning their~purposes.

On the grounds of the results obtained in~\cite{C1} as well as the framework constructed for such, we
can (in view of the previous sections) ponder on the surmise  towards using L\'evy paths in the
context of minisuperspace cosmology. Concretely,  investigating if a fractional quantum cosmology, with~a (consistently) generalized Wheeler--DeWitt equation,  can be obtained.
In more detail, this would mean extending the general path integral (at the basis of the current line conveying fractional quantum mechanics) towards minisuperspace configurations. 
In other words, exploring if the notion of L\'evy paths can be used thereby. The~task of ascending  this summit  would be immense, if~we aim at retrieving a generalized (fractional) version for the Wheeler--DeWitt equation, in~view of the obstacles described in~\cite{C1,C2}. Hence, it would either be this route or, instead, perhaps complementary heuristic insights could be employed, to~
 provide any meaningful results (or at least,  useful bearings)  to guide~us.

In line with the previous paragraph, let us moreover add  the following. In~the procedure to retrieve the Schr\"odinger equation, within~ the Brownian--Feynman path integral (for  standard quantum mechanics), segments as  $x_i(t)$, denoting spatial coordinates in classical time, are used. Whereas a relativistic extension could presume
the use of space-time coordinates $X_\mu(\tau)$, $\tau$ an affine parameter (or just a proper time) for word-line segments in a Minkoswskian framework (assuming no curvature effects);  the symmetries in the former would be merely translations and spatial rotations,  whereas in the latter Lorentzian boosts would be necessary. If~including curvature,  then generic diffeomorphisms would be~required. 




\section{ Fractional Quantum Cosmology: An Heuristic~Approach}
\label{sect-4}

This section
bears twofold content. On~the one hand,  in~Section \ref{new1-subsect} we speculate
how a fractional Wheeler--DeWitt could be written. 
We take, if~we can express ourselves in these terms, a~
'mathematically heuristic' stand: it  advances a 
discussion, declared not to be optimal,  but~which is nevertheless  able to 
bring issues to ponder about. 
On the other hand, in~Section \ref{subsect} we try to be a bit more  savvy. Establishing 
a perfect setting  to investigate is impractical and thus,  heuristic methods are 
instead used to finding a simple case for discussion. These are shortcuts that ease our analysis, we do declare~it.


\subsection{Speculating about a Fractional Wheeler--DeWitt~Equation}
\label{new1-subsect}
 
 An extension of L\'evy paths towards a description of relativistic  space-time (or even a  (mini)superspace) is still quite absent. Therefore, much that can be proposed meanwhile  is entirely heuristic, some in the form of `educated guesses', which is what 
convey in this~subsection.


In most of fractional quantum mechanics,  the~energy operator still merely becomes
$\frac{\partial }{\partial t}$ (as in the usual set up), whereas  the Laplacian instead becomes (see Equations~(\ref{7}) and (\ref{8}))
\begin{equation}
    D_\alpha \left(
    - \hbar^2 \Delta
    \right)^\frac{\alpha}{2}
    \equiv D_\alpha \left(
    - \hbar^2
    \sum_i \frac{\partial^2 }{\partial x_i^2}
    \right)^\frac{\alpha}{2}.
    \label{final-1}
\end{equation}
However, the~Wheeler--DeWitt is (formally) a Klein--Gordon-like equation. In~particular, the~d'Alembertian can be cast (simplified) 
as (It is important to remember that the Schr\"odinger equation  is a variant of the `heat equation' i.e.,~  a parabolic type of PDE, whereas the Klein--Gordon is a  wave equation, an~hyperbolic PDE (which upon Wick rotation can become elliptic); this characteristic is shared by the Wheeler--DeWitt equation for quantum cosmology. This is pertinent, in~terms of proceeding to either extract it from a suitable L\'evy process or, as~we discuss in this section, heuristically build a suitable quantum cosmological framework for that. Concerning the latter, the~nature of the  mathematical PDE types, plus bearing a (classical) Euclidean space or a `relativistic' Lorentzian   signature for minisuperspace is of importance. All this can be relevant when opting to 
discuss  a whole fractional Wheeler--DeWitt equation or, instead,  just a fractional 
Schr\"odinger equation,  bearing  gravitational quantum induced corrections~\cite{CKTP}. In~addition, expression (\ref{final-1}) bears an Euclidean signature, whereas in
(\ref{final-2}) a Lorentzian (Riemannian) manner widens the scope)
, e.g.,
\begin{equation}
\Box \equiv - \frac{\partial^2 }{\partial a^2} + \sum_i \frac{\partial^2 }{\partial \phi_i^2} := {\mathbf{g}}^{ij}(a,\phi_k) \frac{\partial }{\partial q_i} \frac{\partial }{\partial q_j},
\label{final-2}
\end{equation}
 with ${\mathbf{g}}^{ij}$ being a metric for a $(a,\phi_i)$ minisuperspace. The~challenge is that there is yet no relativistic  fractional quantum mechanics formulation. It it would, that could  guide us into better (beyond heuristic or
 just mere speculative) lines towards fractional quantum cosmology. In~particular,  would the Riesz derivative  be placed, too simply, as~\begin{equation}
 \Box_\alpha \equiv - \frac{\partial^2 }{\partial a^2} + D_\alpha \left(
    - \hbar^2
    \sum_i \frac{\partial^2 }{\partial x_i^2}
    \right)^\frac{\alpha}{2}
    \label{final-3a}
\end{equation}
 or instead, still simplified,
\begin{equation}
 \Box_\alpha \equiv D_\alpha  \left(  \hbar^2
 \frac{\partial^2 }{\partial a^2} - \hbar^2 \sum_i \frac{\partial^2 }{\partial \phi_i^2}
 \right)^\frac{\alpha}{2},
    \label{final-3b}
\end{equation}
with $\Box_\alpha$ as generalized d'Alembertian, 
induced from L\'evy path integrals. Perhaps more reasonably, something as
\begin{equation}
 \Box_{\hat \alpha} \equiv
\left( D_{\tilde \alpha}^{(q_{i})} \right)^\frac{1}{2}
\left( D_{\overline\alpha}^{(q_{j})} \right)^\frac{1}{2}
\left[
-\hbar^2
\mathbf{g}^{ij}(q_k)
\frac{\partial}{\partial q_i} \frac{\partial}{\partial q_j}
\right]^\frac{{\hat \alpha}}{2},
\label{final-3ca}
\end{equation}
will be retrieved, with, e.g.,~ $\left\{q_k\right\} \equiv \left\{a,\phi_j\right\}$. Equation~(\ref{final-3ca}) is aiming at matching minisuperspace covariance (see~\cite{CKnew}). Moreover, the~L\'evy index was coined within an Euclidean setting whereas  a (Lorentzian) minisuperspace may now require and `mix' different $\alpha$'s,  per minisuperspace variable, $q_i, q_j$, allowing for the several path components, now in the configuration space, parametrized by  an affine term, e.g.,~ $\tau$. Hence, the~labels $\tilde \alpha$ and $\overline \alpha$, for~$q_i, q_j$, respectively, whereas $\hat \alpha$ symbolically points to  the possibility to allow for this `mixing' and not assuming a unique L\'evy parameter in the more general settings herein. Possibly extending from Equations~(\ref{7}) and (\ref{8}), we could further add, writing for $\Box_\alpha$ in (\ref{final-3ca}), that
\begin{equation}
\Box_{\hat \alpha}  \Psi (q_k; \tau)
=
{\mathcal P}
\int
d^{\mathcal D}\! \pi_k \,
\exp\!\left(i \pi_k q^k / \hbar\right)
|\pi^k|^{\hat \alpha}
\Phi(\pi_k;\tau),
\label{final-3cb}
\end{equation}
with ${\mathcal P}$ a prefactor related to the minisuperpcace dimension ${\mathcal D}$, $\pi_k $ the canonical conjugated momentum to $q^k$; ${\hat \alpha}$ in $|\pi^k|^{\hat \alpha}$ needs to be specified, as~related to the possible range of $\alpha$'s allowed but it may be that a sole L\'evy parameter is~ever-present.

The 
issue of a fractional time derivative brought into the Schr\"odinger equation is of interest to mention at this 
point. In~fact 
(see~\cite{TIME} and the many references therein on the issue), fractional time derivatives have been considered, allowing to discuss issues such as non-unitarity and strictly taking a canonical approach.  Extending the framework of L\'evy paths and fractional (quantum) mechanics to relativistic settings, it could impose  to adequately import the results 
from the explorations in \cite{TIME} and alike. However, if bearing  intrinsic minisuperspace covariance \cite{CKnew}, would then unitarity be regained? Let us just mention that non-unitarity also emerges in discussions about semiclassical quantum gravity~\cite{CKTP}, namely from a Wheeler--DeWitt expansion towards obtaining a Schr\"odinger equation (as well a WKB-like "many-fingered" functional time) in the presence of quantum gravitational corrections, whereby that covariance is~lost.



\subsection{Fractional Quantum FLRW Cosmology:\\ A Simple Case~Study}
\label{subsect}

As a simple toy model, let us consider a Friedmann-Lema\^{i}tre-Robertson-Walker (FLRW) universe with the following line element
\begin{eqnarray}\label{s1}
ds^2=-N^2(t)dt^2+a^2(t)\Big[\frac{dr^2}{1-kr^2}+r^2d\Omega^2\Big],
\end{eqnarray}
where $N(t)$ is the lapse function, $a(t)$ is the scale factor and $k=\pm1,0$ represents the spatial 3-curvature of a homogeneous and isotropic 3-dimensional (compact and without boundary) hypersurface, $\Sigma_t$.
The compactness of universe indicate that its 3-volume $\mathcal V_k$ is finite. The~ADM action functional of the gravitational part plus matter fields (herein a perfect fluid with energy density $\rho$) is~\cite{E}
\begin{eqnarray}\label{S2}
S=\frac{1}{16\pi G}\int_{t_i}^{t_f}dt\int_{\Sigma_t}d^3xN\sqrt{h}\Big(^{(3)}R+K_{ij}K^{ij}-K^2\Big)-\int_{t_i}^{t_f}dt\int_{\Sigma_t}d^3xN\sqrt{h}\rho,
\end{eqnarray}
where $^{(3)}R$, $K_{ij}$ and $h_{ij}$ are the Ricci scalar, the~extrinsic curvature and the induced metric of $\Sigma_t$ respectively. For~this FLRW universe, the~action will simplify to
\begin{eqnarray}\label{S3}
S=\frac{3\mathcal V_k}{8\pi G}\int_{t_i}^{t_f}\Big(-\frac{a\dot a^2}{N}+kNa\Big)-\mathcal V_k\int_{t_i}^{t_f} Na^3\rho dt,
\end{eqnarray}
where a overdot denotes differentiation with respect time coordinate $t$.
Let us assume the matter content of universe is \textit{non}-interacting dust 
and radiation, i.e.,~ $\rho=\rho_\gamma+\rho_d$, where $\rho_\gamma$ and $\rho_d$ are the corresponding energy density of radiation and dust respectively. The~conservation of the perfect fluids $\rho_\gamma$ and $\rho_d$ leads to $\rho_\gamma=\rho_{\gamma0}(a/a_0)^{-4}$ and $\rho_d=\rho_{d0}(a/a_0)^{-3}$,  where $a_0$ and $\rho_{0}$ are the values of the scale factor and the energy density of a fluid,  at~a measurement epoch $t_0$.
By using the following definitions
\begin{eqnarray}\label{S4}
\begin{array}{ccc}
  \Omega_{0,\gamma}:=\frac{8\pi G\rho_{\gamma0}}{3H_0^2},\,\,\,\,  \Omega_{0,d}:=\frac{8\pi G\rho_{d0}}{3H_0^2},\,\,\,\,\Omega_{0,k}:=-\frac{k}{a_0^2H_0^2},\\
  \tilde N(t):=\frac{N(t)}{H_0x(t)},\,\,\,x(t):=\frac{a}{a_0}+\frac{\Omega_{0,d}}{2\Omega_{0,k}},\,\,\,\,d\eta:=H_0dt,\,\,\,\,M:=\frac{3\mathcal V_ka_0^3H_0}{8\pi G},
\end{array}
\end{eqnarray}
where $H_0$ is the Hubble parameter at the measurement time $t_0$,  action (\ref{S3}) 
further simplifies
to
\begin{eqnarray}\label{S5}
S=-\frac{M}{2}\int_{t_i}^{t_f}\Big(\frac{\dot x^2}{\tilde N}+\tilde N\Big(\Omega_{0,k}x^2+\Omega_\gamma-\frac{\Omega_{0,d}^2}{4\Omega_{0,k}}\Big)\Big)d\eta,
\end{eqnarray}
where now an over-dot denotes differentiation respect to a new time coordinate,  $\eta$. Note that all density parameters $\Omega_{0,i}$ defined in (\ref{S4}) are constants and their values are associated with a measurement time, say $t_0$.
The Hamiltonian constraint is
\begin{eqnarray}\label{S6}
\mathcal H=\tilde N\Big[-\frac{p^2}{2M}+\frac{1}{2}M\Omega_{0,k}x^2+\frac{M}{2}\Big(\Omega_{0,\gamma}-\frac{\Omega^2_{0,d}}{4\Omega_k}\Big)\Big]\approx0,
\end{eqnarray}
where $p=-\frac{M}{\tilde N}\dot x$ is the conjugate momenta of scale factor $x$. At~
$t_0$ the above Hamiltonian constraint gives us the following well-known relation between density parameters
\begin{eqnarray}\label{S7}
\Omega_{0,\gamma}+\Omega_{0,d}+\Omega_{0,k}=1.
\end{eqnarray}

In order to have a setting to comparatively appraise, we now elaborate on the model herein but yet without any fractional calculus (induced) features. I.e., it will be standard quantum~cosmology. 

In the coordinate representation $\hat p:=-i\hbar d/dx$ $\hat x:=x$, the~WDW equation is retrieved as
\begin{eqnarray}\label{S8}
-\frac{\hbar^2}{2M}\frac{d^2\psi(x)}{dx^2}+\frac{1}{2}M\omega^2x^2\psi(x)=\frac{M}{2}\left(\Omega_{0,\gamma}-\frac{\Omega^2_{0,d}}{4\Omega_{0,k}}\right)\psi(x),
\end{eqnarray}
where $\omega^2:=-\Omega_{0,k}=\frac{1}{H_0^2a_0^2}$.
Let us investigate the closed universe (positive sectional curvature) where $k=1$; for more details, see~\cite{E}. In~this case, $\Sigma_t=\mathbb{S}^3/\Gamma$ where $\Gamma$ is the discrete subgroups of $SO(4)$ without fixed point, acting freely
and discontinuously on $\mathbb{S}^3$. Hence, $\mathcal V_{k=1}=\frac{2\pi^2}{|\Gamma|}$,  where $|\Gamma|$ is the order of the group $\Gamma$. For~topologically complicated spherical 3-manifolds, $|\Gamma|$ becomes large and consequently the volume is small, $0<\mathcal V_{k=1}\leq2\pi^2$. There is no lower bound since $\Gamma$ can have an arbitrarily large number of~elements.

We further note that the domain of definition of the scale factor is $x\in {\mathbb R}^+$. Consequently, the~operator $H:=-\frac{\hbar^2}{2M}\frac{d^2}{dx^2}+\frac{1}{2}M\omega^2x^2$ in the left hand side of (\ref{S8}) is defined on a dense domain $C^\infty(\mathbb R^+)$ and it is in the limit point case at $+\infty$ and in the limit circle case at $x=0$. Hence, $H$ is not essentially a self-adjoint operator. It constitutes a  symmetric  Hermitian  operator if
\begin{eqnarray}\label{SS1}
\langle\psi_1|H\psi_2\rangle=\langle H\psi_1|\psi_2\rangle, \,\,\,\,\psi_1,\psi_2\in {\mathcal D}(H),
\end{eqnarray}
or equivalently
\begin{eqnarray}\label{SS2}
\lim_{x\rightarrow 0^+} \left(\frac{d\psi^*_1}{dx}\psi_2-\psi^*_1\frac{d\psi_2}{dx}\right)=0.
\end{eqnarray}
To guarantee the validity of this condition, it is necessary and sufficient that
\begin{eqnarray}\label{SS3}
\left(\frac{d\psi(x)}{dx}+\gamma \psi(x)\right)_{x=0^+}=0,\,\,\,\,\forall \psi(x)\in {\mathcal D}(H),
\end{eqnarray}
where $\gamma$ is an arbitrary real constant. This shows that the parameter $\gamma$ characterize a one-parameter family of self-adjoint extensions of $H$. The~general square-integrable solution of Equation~(\ref{S8}) is
\begin{eqnarray}\label{SS4}
\begin{array}{cc}
\psi(x)=\sqrt{\pi}e^{-\frac{M\omega}{2}x^2}\,_1F_1\left(\frac{1}{4}-\frac{E}{2\omega};\frac{1}{2};\frac{M\omega}{2}x^2\right)
\\-\frac{\sqrt{\pi M\omega}xe^{-\frac{M\omega}{2}x^2}2^{\frac{3}{4}\frac{E}{2\omega}}}{\Gamma\left(\frac{1}{4}-\frac{E}{2\omega}\right)}\,_1F_1\left(\frac{3}{4}-\frac{E}{2\omega};\frac{3}{2};\frac{M\omega}{2}x^2\right),
\end{array}
\end{eqnarray}
where $E:=\frac{M}{2}\left(\Omega_\gamma-\frac{\Omega^2_d}{4\Omega_k}\right)$, $\Gamma(a)$ is Gamma function and $_1F_1(a;b;x)$ is confluent hypergeometric function. By~using the properties $_1F_1(a;b;0)=1$ and $\frac{d}{dx}\,_1F_1(a;b;x)=\frac{a}{b}\,_1F_1(a+1;b+1;x)$, we can rewrite the Robin boundary condition (i.e., expression (\ref{SS3})) as
\begin{eqnarray}\label{SS5}
\gamma=2\sqrt{M\omega}\frac{\Gamma(\frac{3}{4}-\frac{E}{2\omega})}{\Gamma(\frac{1}{4}-\frac{E}{2\omega})}.
\end{eqnarray}
Regarding that the parameter $\gamma$ has dimension of inverse of length (as pointed out in~\cite{Tipler}), then $\gamma$ would be a new fundamental constant of theory. However,~as addressed  in~\cite{Jalalzadeh}, the~origin of this unwanted new constant is the effective matter field Lagrangian in action (\ref{S2}). If~we use a ``real'' matter field, for~example a scalar field or the Maxwell's field Lagrangian instead of $\rho$ in (\ref{S2}), then the value of $\gamma$ will be fixed to only following two acceptable values
\begin{eqnarray}\label{SS6}
\gamma=0,\,\,\,\,\,\textrm{or}\,\,\,\,\,\frac{1}{\gamma}=0.
\end{eqnarray}
Using these values of $\gamma$, we obtain
simple harmonic oscillator states,  with~eigenvalues
\begin{eqnarray}\label{SS7}
E=\hbar\omega\left(n+\frac{1}{2}\right),
\end{eqnarray}
where $n$ is an even or odd integer corresponding to the first or the second value of $\gamma$ in (\ref{SS6}), respectively.
The relation (\ref{SS7}) gives us the eigenvalues of WDW Equation~(\ref{S8})
\begin{eqnarray}\label{S9}
\frac{\kappa}{2\omega^3}\left(\Omega_{0,\gamma}-\frac{\Omega^2_{0,d}}{4\omega^2}\right)=\hbar\omega\left(n+\frac{1}{2}\right),
\end{eqnarray}
where $\kappa:=\frac{3\mathcal V_k}{8\pi GH_0^2}$. For~large values of the quantum number $n$ (or very small values of $\kappa$) and also for finite values of density parameters $\Omega_{0,\gamma}$, $\Omega_{0,d}$ and $\Omega_{0,k}$, the~above eigenvalue relation will reduce to the following three relations
\begin{eqnarray}\label{S10}
\begin{array}{cc}
\Omega_{0,k}\simeq-\frac{1}{2}\left(\frac{\kappa}{\hbar (n+\frac{1}{2})}\right)^\frac{1}{2},\\
\\
\Omega_{0,d}\simeq\left(\frac{2\kappa}{\hbar (n+\frac{1}{2})}\right)^\frac{1}{4},\\
\\
\Omega_{0,\gamma}\simeq1.
\end{array}
\end{eqnarray}
If we assume the universe has displayed (\textit{circa} its  beginning) a grand unified setting, by~$t_0\simeq10^{-43}$s~\cite{Allday,Allday-n},  following the Planck epoch, $t_{Pl}\simeq10^{-44}$s, then
\begin{eqnarray}\label{S11}
\kappa=\frac{3\mathcal V_k}{8\pi GH_0^2}\simeq\frac{15\hbar\pi}{2|\Gamma|}.
\end{eqnarray}
Therefore, at~the beginning of a grand unified theory dominance, 
the values of density parameter $\Omega_k$ and $\Omega_d$ will reduce to
\begin{eqnarray}\label{S12}
\Omega_{0,k}\simeq-\left(\frac{15\pi}{8|\Gamma|(n+\frac{1}{2})}\right)^\frac{1}{2},\,\,\,\,\Omega_{0,d}\simeq\left(\frac{15\pi}{|\Gamma|(n+\frac{1}{2})}\right)^\frac{1}{4},\,\,\,\,\Omega_{0,\gamma}\simeq1.
\end{eqnarray}
These relations show that for a large value of quantum number $n$ (or for a complicated geometry, $\mathbb{S}^3/\Gamma$), the~emerged classical universe will be very close to spatially flat and radiation~dominated.

\begin{figure}[h]
    \centering
    \includegraphics[width=0.5\textwidth]{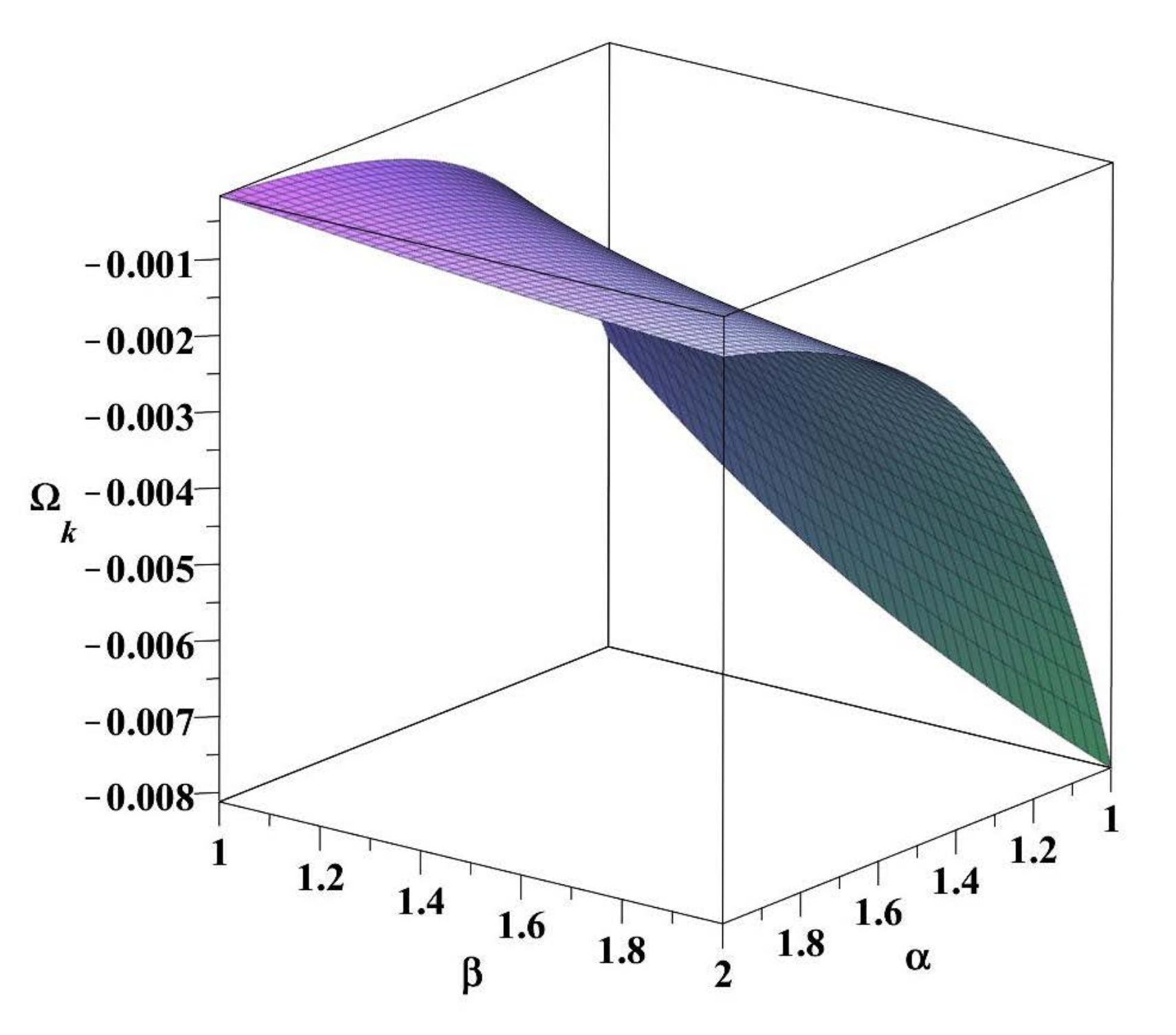}
    \caption{The plot of $\Omega_{0,k=1}$ for $n=500000$ and $|\Gamma|=1$.}
    \label{fig}
\end{figure}

We now study the fractional quantum cosmology of the model. Following  Equation~(\ref{9}), an~applicable (and simplified to be workable) 
fractional version of the Wheeler--DeWitt Equation~(\ref{S8}) for $k=1$ will be
\begin{eqnarray}\label{S13}
-\frac{M}{2}\left(\frac{\hbar}{M}\right)^\alpha\frac{d^\alpha\psi}{dx^\alpha}+\frac{1}{2}M\omega^2x^\beta\psi=\frac{M}{2}\left(\Omega_{0,\gamma}+\frac{\Omega_{0,d}}{4\omega^2}\right)\psi,
\end{eqnarray}
where $\omega^2=-\Omega_{0,k}$. Moreover,  following~\cite{Herrmann}, we assumed that $D^\alpha:=\frac{M}{2}\left(\frac{\hbar}{M}\right)^\alpha$. The~semiclassical eigenvalue of this equation has already been obtained in Section \ref{HO}. So, relation (\ref{13b}) gives us
\begin{eqnarray}\label{S14}
\Omega_{0,\gamma}+\frac{\Omega^2_{0,d}}{4\omega^2}\simeq\left(\frac{\pi\hbar\beta(n+\frac{1}{2})}{\kappa B(\frac{1}{\beta},\frac{1}{\alpha}+1)}\right)^\frac{\alpha\beta}{\alpha+\beta}\omega^\frac{3\alpha\beta+3\beta-\alpha}{\alpha+\beta}.
\end{eqnarray}
Again, for~$\alpha=\beta=2$, we recover (\ref{S9}).
Therefore, for~finite values of density parameters and large values of quantum number $n$ (or for complicated geometries) at the beginning of grand unified theory the values of density parameters for fractional quantum cosmology will be
\begin{eqnarray}\label{S15}
\begin{array}{ccc}
\Omega_{0,k}\simeq-\left(\frac{15\pi B(\frac{1}{\beta},\frac{1}{\alpha}+1)}{2^\frac{\alpha+\beta+\alpha\beta}{\alpha+\beta}\beta|\Gamma|(n+\frac{1}{2})}\right)^\frac{2\alpha\beta}{2\alpha\beta+3\beta-\alpha}\rightarrow0,\\
\\
\Omega_{0,d}\simeq\sqrt{-2\Omega_{0,k}}\rightarrow0,\\
\\
\Omega_{0,\gamma}\simeq1.
\end{array}
\end{eqnarray}
 Figure~(\ref{fig}) shows the graph of $\Omega_{0,k}$ for $n=500000$ and $|\Gamma|=1$, as~an example. It shows an interesting feature of the fractional quantum cosmology of this simple model: the smaller values $\alpha\rightarrow2$ and $\beta\rightarrow1$, give us the smaller values for the density parameters of sectional curvature $\Omega_{0,k}\rightarrow-0.0002$ and dust $\Omega_{0,d}\rightarrow 0.02$.

We could conclude that with fractional quantum cosmology we have a powerful tool to control and maybe remove the flatness problem of standard cosmology,  without~any need to invoke the inflation paradigm.

\section{Discussion  and~Outlook}
\label{sec-5}


Let us summarize and close this paper by mentioning the~following. 

By means of this merely introductory paper, we 
presented herewith a set of heuristic ideas that will, surely, constitute motivation and  enthuse more  
work on a compelling subject, which we hold as  enticing. We stress
  that,  isolated, those ideas (cf. Sections~\ref{sec-1}--\ref{sec-3}) exist elsewhere  in the literature~\cite{A,B,C1,C2}; embracing them altogether now (cf. Section~\ref{sect-4}) is the
 advancing step we bring  herewith.
Further elements and features beyond  the explicit 
content in Section~\ref{sect-4}
are postponed to e.g.,~either~\cite{E} or   forthcoming publications, hopefully   
from other authors. An~allegory for the overture we convey  is that of an unlocked window disclosing  a potential fruitful but trying landscape,  rather than displaying an (albeit new) orderly preset  ground, where to  quickly cultivate fine-tuned seeds and cleanly harvest from them. Reiterating, 
either from the Abstract or the Introduction hereby, the~objective of this manuscript was solely to  bequeath to its readers a set of probationary  lines, sometimes implicitly in the text, for~future assessment within the eventual construction of a fractional quantum~cosmology.

However, fairness dictates that we emphasize that  Section~\ref{sect-4} indeed risks the surmise of bearing few specific  claims, which, we keep pointing out,  may nevertheless   prove  worthwhile into questioning and improve the current state of affairs. If~this occurs, then the aim of advancing and following from  the title of this paper will be satisfied. Indeed, much more and significant progress is needed. Notwithstanding the eagerness of this paper purpose, 
 we   advocate meanwhile a few tentative discussions,  tempered with adequate reserve. A~few  lines to consider for investigation  would be as follows: 

\begin{enumerate}
    \item To begin with, let us recall (cf. Section~\ref{sec-2}) that implementing a viewpoint and methodological change, namely from  Brownian towards L\'evy paths,  conducted  to
fractional quantum mechanics~\cite{A,B}. Interesting applications include the (harmonic) oscillator, particular cases of tunneling and the Hydrogen atom. Employing  L\'evy paths into quantum cosmology would lead to a fractional Wheeler--DeWitt equation, we conjecture. In~other words,  to~generalize straight from~\cite{C1}
but within L\'evy paths. This may bring us a
far more robust and mathematical coherent (generalized) fractional Wheeler--DeWitt~equation.

A serious (and also  important) issue to explore and settle would be  about minisuperspace covariance~\cite{CKnew}, explicit within the  d'Alembertian as pointed out in textbooks of quantum cosmology~\cite{C2,CKnew}. However,~within fractional quantum cosmology,    extracted from L\'evy paths, would a generalized Wheeler--DeWitt equation maintain it, alter it or eliminate it? Moreover, a~fractional Wheeler--DeWitt (likewise for the Schr\"odinger) equation could be be an integro-differential equation. Besides~complicated analytical considerations, numerical ingredients and analysis would be~mandatory.

    \item When retrieving the Schr\"odinger equation from  the Brownian--Feynman path integral (i.e., standard quantum mechanics), segments as $x_i(t)$, representing classical spatial coordinates in classical time, are considered.  For~a  relativistic extension,   space-time coordinates $X_\mu(\tau)$  ($\tau$ possibly just a proper time) for word-line segments could 
    be contemplated. This step is yet to be attempted (to these author's knowledge) in fractional quantum mechanics, namely bring it within L\'evy paths~\cite{B,B-fqm,NEW}. Only then (mini)superspace configurations 
    would be properly discussed, bearing some of Section \ref{subsect-levy} features~\cite{B,NEW}; any extension of L\'evy paths towards a description of curved space-time (or instead a (mini)superspace) is still absent. A~ fair contribution towards a rigorous description is needed, to~proceed beyond heuristic~appraisals. 
    
    \item From a strict, purely mathematical point of view, the~use of dimensions  within  formulae  with physical  observables  is  meaningless.   However,~ if  proceeding towards a physical system, this issue could become of importance. When bringing fractional derivatives, how will  realistic tests and data comparison be done? There are publications discussing it or at least, the~mathematical-physical framework. Simplistically,  as~pointed out, could the physical (i.e., dimensional) consequences of using fractional derivatives become hidden in constants, taken as parameters to fit~\cite{Cresson,Inizan}?   
    
    \item Related to  the above item and as we have mentioned, 
    fractional quantum mechanics has not 
    been taken and discussed concerning  experimentation, even if just for  \textit{Gedankenexperiment}. It is  not yet reachable with the current technology. 
    However, let us revisit  the  discussions about effective mass in concrete Bose-Einstein condensates~\cite{Pin}. Nevertheless, no  experimental lines have provided any guidance concerning any of the parameters,  such as the L\'evy index, etc. However,~cf. references~\cite{Pin,Pin-a}.  
    Fractional quantum mechanics has not yet been tested, though~ it is falsifiable. `Situation room': fairness  points that it is, to~this age, consistent, in~that includes standard quantum mechanics (as clear limiting cases through parameter variation). 
    
    So, would fractional quantum cosmology be able to  provide predictions  that would prove observational inconsistent or narrowed for consistency? Significant more work  is needed to achieve that~stage.
    
    
    \item Albeit  working on a rather simplified FLRW cosmological model, we  envisaged how fractional calculus induced elements  (imported to some judicious extent) could change very specific features. Namely, the~discussion on a particular application within a  FLRW model:  it allowed   to speculate on  the flatness problem of standard cosmology. We are aware of the perhaps uncomplicated  assumptions we took in employing therein fractional calculus. Proceeding into more rigorous mathematical computations may require to use far more elaborated expressions, possibly not even those in (\ref{final-3a})--(\ref{final-3cb}) but other improved~formulae.

      Other issues, such as the horizon or structure formation should of course be considered. This surely must and will be discussed in subsequent publications, possibly by other authors whom we challenge to contribute as well. Likewise, other broader cosmologies or matter contents would be important to investigate:   more should be done towards appraising fractional quantum~cosmology. 
    
    \item A paradigmatic setting in quantum cosmology has been the
(harmonic) oscillator, used,   within~the FRW one-dimensional minisuperspace
models, with~$H= p_a^2 + U(a)$, $U(a) \sim  a^2$, then $U_\Lambda(a) \sim  a^2 - \Lambda a^4  $; cf. solutions
as DeSitter and conformally coupled scalar field minisuperspace~\cite{C2}. 

A rather specific issue, relevant for quantum cosmological applications,  would be to have
fractional quantum mechanics further explored and  elaborated, particularly
concerning tunneling within a  WKB approach for e.g.,~potentials of the form  $U_\Lambda(a) \sim  a^2 - \Lambda a^4  $. Being more concrete, exploring the situation (with either $E=0$ or $E\neq 0$) of nucleation from classical forbidden to allowed regions or a transition from classical allowed, through classical forbidden, towards classically allowed domains. Within~quantum cosmology, tunneling (nucleation) is usually taken with $E=0$ (as following from the $H=0$ constraint) but a $E_{rad} \neq 0$ has been explored in~\cite{Alwis,MBL-PVM} (cf. references therein, too), within~concrete applications for the wave function of the universe and (initial) conditions.

Furthermore,  since variance can emerge as  asymptotic infinite in  fractional quantum mechanics
~\cite{B,B-fqm,T1,T2},
then processes could be more likely to occur (or not) regarding Universe nucleation,  initial conditions for inflation and its  likelihood  within standard 
    quantum cosmology \textit{versus} a fractional framework. This modified setting could be explored  with respect to sampling initial~conditions.

    \item Finally, simply take and `play', aiming to induce  a fractional (quantum gravitational modified) Schr\"odinger equation, within~the principles present in~\cite{CKTP}. We could simply
directly
modify, \textit{ad-hoc},  the~Laplacian therein
(in the Schr\"odinger equation) to  further probe it. For~instance, about  the Bunch--Davies state
or a deviation, now within a fractional quantum mechanics setting. Or~instead about non-unitarity following from the quantum gravitational corrected Schr\"odinger equation~\cite{CKTP}; could that non-unitarity  be re-cast as a consequence either of minisuperspace covariance being lost or, \textit{equivalently},  fractional time derivative emerging? 

It would be thus  immensely interesting if a Schr\"odinger equation bearing gravitational quantum induced corrections~\cite{CKTP}, but~within fractional quantum mechanics, could be investigated, eventually applied to concrete cases. In~addition, discussing whether the seeding process from fluctuations in a scalar field $\delta \phi$ would be `easier' to emerge cf.~\cite{T1,T2}). Or~would the  Bunch--Davies vacuum,  associated with Gaussian states and a Schr\"odinger equation for matter fields (when dealing with quantum gravitational corrections~\cite{C2,CKnew,CKTP}) be removed and other quite different state     be retrieved (by means of  a suitable fractional Schr\"odinger equation)?     
    
\end{enumerate}




In addition, there is plenty still to be done. 
Thus, remembering and respectfully borrowing  from Leibnitz~\cite{A},  ``($\dots$)  useful consequences will be~drawn''. 
 
%
%

\vspace{6pt}

\section*{Acknowledgement}{All authors contributed equally.}
{This research work was supported by Grant No. UID/MAT/00212/2019,  COST Action CA15117 (CANTATA) and COST Action CA18108 (Quantum gravity phenomenology in the multi-messenger approach).} 

{The authors are grateful to Jos\'e Velhinho for his kind invitation to write this paper. We thank the referees for their  constructive remarks, assisting us
in producing a manuscript that    in earnest 
may  entice for subsequent work.   
PVM acknowledges  DAMTP and Clare Hall, Cambridge for kind hospitality and
a Visiting Fellowship during his sabbatical, during~which this work was initiated. }



\end{document}